\documentclass[superscriptaddress,floatfix,reprint]{revtex4-1}
\usepackage[latin1]{inputenc}
\usepackage[american]{babel}
\usepackage{graphicx}							
\usepackage{dcolumn}
\usepackage{bm}
\usepackage{here}
\usepackage{amsmath,amssymb}     
\usepackage{color}

\begin{document}		

\author{Alexander Hinderhofer} \affiliation{Institute for Applied Physics, University of T\"ubingen, Auf der Morgenstelle 10, 72076 T\"ubingen, Germany}
\author{Jan Hagenlocher} \affiliation{Institute for Applied Physics, University of T\"ubingen, Auf der Morgenstelle 10, 72076 T\"ubingen, Germany}
\author{Alexander Gerlach} \affiliation{Institute for Applied Physics, University of T\"ubingen, Auf der Morgenstelle 10, 72076 T\"ubingen, Germany}
\author{Joachim Krug} \affiliation{Institute for Biological Physics, University of Cologne, Z\"ulpicher Strasse 77, 50937 K\"oln, Germany}
\author{Martin Oettel} \affiliation{Institute for Applied Physics, University of T\"ubingen, Auf der Morgenstelle 10, 72076 T\"ubingen, Germany}
\author{Frank Schreiber} \affiliation{Institute for Applied Physics, University of T\"ubingen, Auf der Morgenstelle 10, 72076 T\"ubingen, Germany}

\date{\today}
\title{Non-equilibrium Roughness Evolution of Small Molecule Mixed Films Reflecting Equilibrium Phase Behavior}

\begin{abstract}
Understanding non-equilibrium phenomena, such as growth, and connecting them to equilibrium phase behavior is a major challenge, in particular for complex multicomponent materials.
We use X-ray reflectivity to determine the surface roughness of binary mixtures of several prototypical organic compounds.
By analyzing the roughness as a function of composition, we find a systematic behavior depending on the bulk phase behavior in terms of intermixing, co-crystallization or phase separation. 
Supported by kinetic Monte Carlo simulations, we provide evidence that the growth behavior can be rationalized by a lowered step edge barrier in the mixed films which is induced by reduced in-plane crystallinity.

\end{abstract}

\maketitle


\section{Introduction}
Structure formation involves intriguing aspects of non-equilibrium statistical physics. Complexity arises from the fact that in contrast to equilibrium thermodynamics, it is not sufficient to determine only the lowest free energy state, although the equilibrium scenario is expected as a limiting case.
A typical example for a non-equilibrium system is thin film growth, where a large variety of morphologies can be observed, which cannot be explained solely by equilibrium considerations. Instead, kinetic effects play a strong role in the formation of microscopic and mesoscopic structures. A key observable to characterize the growth behavior is the surface roughness ($\sigma$, standard deviation of the film thickness), which is also of substantial technological importance \cite{Barabasi_1995_book,Pimpinelli_1998_book,TMichely_2004_book,Krug_1997_AdvPhys,Lu_2001_book}.

The growth of molecular thin films was studied extensively both theoretically \cite{Pimpinelli_2014_JPhysChemLett,Hlawacek_2008_Science,Hlawacek_2013_JPhysCondensMatter} and experimentally \cite{Hinderhofer_2010_EurophysLett,Yim_2009_ApplPhysLett,Yang_2015_SciRep,Zhang_2010_JPhysChemC,Nahm_2017_JPhysChemC,Hinderhofer_2012_ApplPhysLett}.
Frequently, a typical feature for crystalline molecular films is the comparably fast roughening, often expressed as a high roughening exponent \cite{Durr_2003_PhysRevLett,Zhang_2010_JPhysChemC,Storzer_2017_JPhysChemC}. The complex roughening behavior is explained predominantly by kinetic effects based on high step edge barriers \cite{Fendrich_2007_PhysRevB,Roscioni_2018_JPhysChemLett,Goose_2010_PhysRevB,Ehrlich_1966_JChemPhys,Schwoebel_1966_JApplPhys}, thickness dependent strain release \cite{Durr_2003_PhysRevLett} or a restricted diffusion length due to defects or grain boundaries \cite{Winkler_2016_}.

A still more challenging case for complex materials are binary molecular systems, which are also important due to their electronic properties. Molecular mixed thin films are studied both with small mixing ratios ($\approx$ 1:100) \cite{Conrad_2008_PhysRevB,Kleemann_2012_OrgElectron,Schwarze_2016_Science} for doping as well as large mixing ratios ($\approx$ 1:1) for bulk heterojunctions and molecular complex formation \cite{Zhang_2017_AccChemRes,Hinderhofer_2012_ChemPhysChem,Aufderheide_2012_PhysRevLett,Dieterle_2015_JPhysChemC,Dieterle_2017_PhysStatusSolidiRRL,Belova_2017_JAmChemSoc,Reinhardt_2012_JPhysChemC,Broch_2013_JChemPhys,Hinderhofer_2011_JChemPhys}.
Dependent on the effective interactions of the compounds, binary systems exhibit several different mixing behaviors in the bulk, such as solid solution, co-crystallization or phase separation \cite{Kitaigorodsky_1984_book,Hinderhofer_2012_ChemPhysChem}.
The structure and morphology resulting from the growth, including the distribution of the two components A and B, strongly impact the effective electronic and optical properties and thus ultimately device performance. 
From the growth perspective, the relationship between mixing behavior in equilibrium and kinetically determined surface roughness is of significant fundamental interest. 

Here, we demonstrate that the mixing ratio and bulk phase behavior correlate strongly with kinetically limited growth effects and specifically with the roughness evolution.
We provide a comprehensive study of a broad range of blends with different electronic and steric characteristics. 
Supported by kinetic Monte Carlo (KMC) simulations we identify three main effects:
1) A general smoothing effect for mixed films is induced by a lowered step edge barrier compared to the pure films.
2) Mixtures forming a co-crystal exhibit a local roughness maximum at 1:1 mixing ratio, because pure phase systems exhibit an increased step edge barrier compared to random mixtures.
3) Strongly phase separating mixtures exhibit increased roughness due to 3D island growth on a larger lateral scale.

\begin{figure}
	\begin{center}
		\includegraphics[width=4cm]{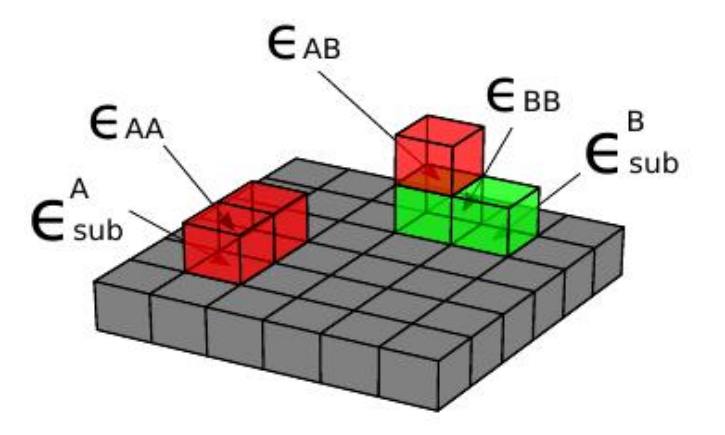}
		\caption{Energy parameters in the binary lattice gas model.}
		\label{fig:energetics}
	\end{center}
\end{figure}

\begin{figure*} [ht]
	\begin{center}
	\includegraphics [width=15.5cm] {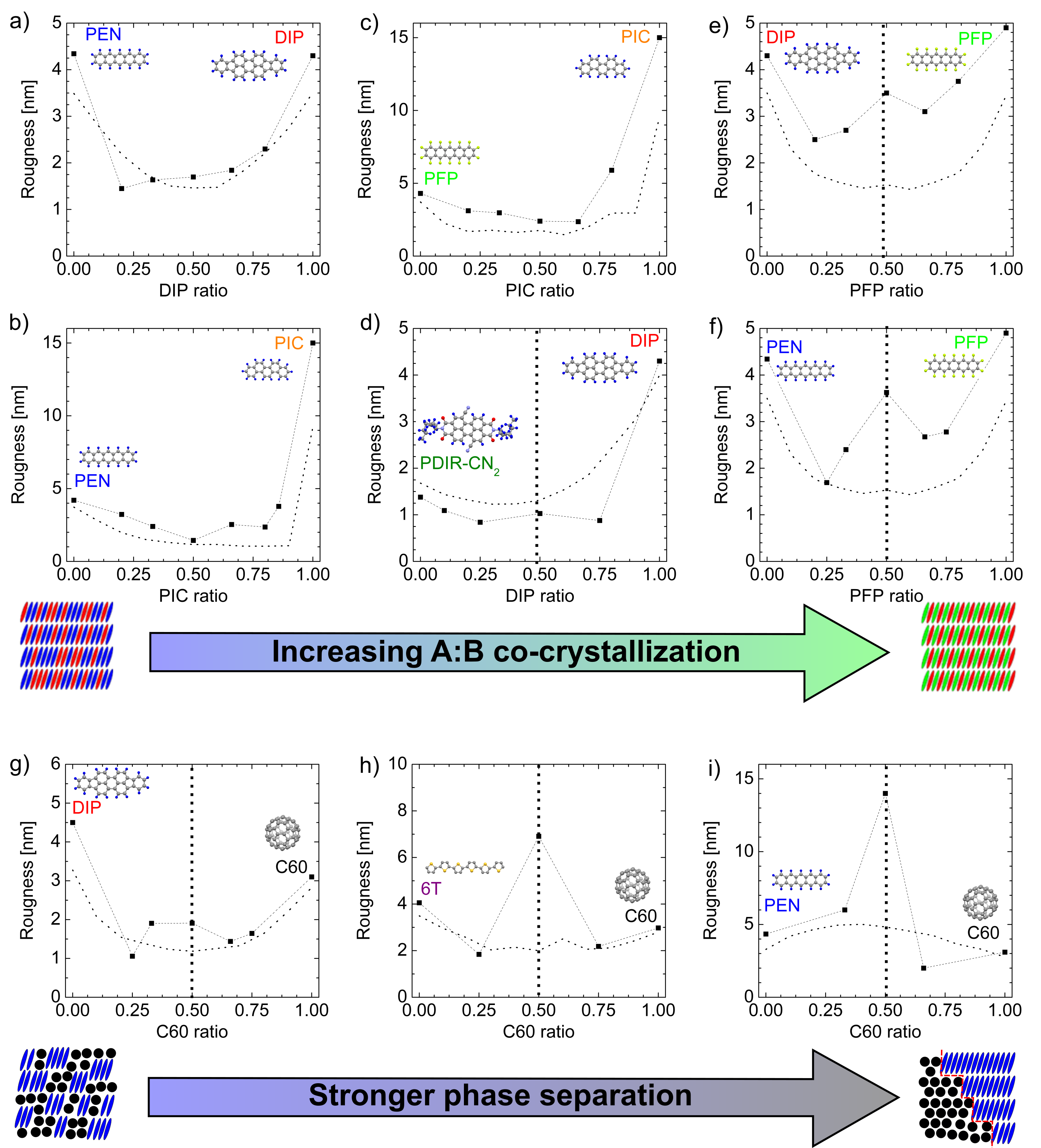}
	\caption{Roughness $\sigma$ of mixed films (20\,nm thickness) with rod like compounds (a-f) dependent on mixing ratio: a) PEN:DIP\cite{Aufderheide_2012_PhysRevLett} b) PEN:PIC\cite{Dieterle_2015_JPhysChemC,Dieterle_2017_PhysStatusSolidiRRL} c) PFP:PIC\cite{Dieterle_2015_JPhysChemC} d) DIP:PDIR-CN$_2$\cite{Belova_2017_JAmChemSoc} e) DIP:PFP\cite{Reinhardt_2012_JPhysChemC,Broch_2013_JChemPhys} f) PEN:PFP\cite{Hinderhofer_2011_JChemPhys}. From a)-f) the in-plane co-crystallization is increasing, i.e.\ PEN:DIP are nearly randomly intermixing, whereas PEN:PFP exhibit well ordered co-crystallization and b)-e) are intermediate cases. (g-i) shows the roughness $\sigma$ of phase separating compounds: g) DIP:C60 h) 6T:C60 i) PEN:C60. The degree of phase separation is increasing from g) to i). Local roughness maxima at ratio 0.5 are marked by vertical lines. All $\sigma$ values were determined by XRR, except those for pure PIC and mixtures of PEN:C60 and 6T:C60, which were determined by AFM. Simulated roughness values are shown as dotted lines. Simulation parameters are listed in supplementary Table 3.}
	\label{lab:fig:rough}
	\end{center}
\end{figure*}

The experiments are performed with a rich variety of molecular species, differing in shape and interaction anisotropy. 
In order to identify the generic behavior, in our simulation approach we model the two--component systems generically with a simple binary lattice gas 
(species A and B). The static parameters governing the equilibrium phase behavior are given by
nearest-neighbor interaction energies $\epsilon_{ij}$ $(i,j=\{\text{A,B}\})$ and substrate interaction energies 
$\epsilon_\text{sub}^\text{A}$ and $\epsilon_\text{sub}^\text{B}$ for particles in the first layer (see Fig.~\ref{fig:energetics}).
Film growth is modeled with KMC simulations with solid--on--solid restrictions (see Sec.~\ref{sec:sim_model}), and the associated dynamic parameters are: (i) free diffusion constants
$D_\text{A[B]}$ for a particle of species A[B] which is not laterally bound, (ii) a deposition rate $F$ (particles per unit of time
and lattice site) and (iii) species--dependent Ehrlich--Schwoebel barriers $E^\text{ES}_{ij}$.
(An important dimensionless ratio determining the degree of nonequilibrium is given by $\Gamma=D_\text{A}/F$.)
All energetic parameters are given in units of the thermal energy.

We stress that we use the simple solid--on--solid model mainly as a conceptual tool. The model is clearly not material--specific, nor is it intended to faithfully represent the microscopic molecular moves and their associated rates. We also work with lower interaction energies and higher deposition rates than in the experiment. However, the simplicity of the model allows to uncover and quantify the three generic effects described above, and to exclude other possible sources of the observed roughness behavior. As detailed molecular simulations of binary growth systems with realistic parameters are currently out of reach, this strategy appears to be most appropriate for elucidating the universal patterns that we see.

	In the following we will first (Secs.~~\ref{sec:rough_intermixing}--\ref{sec:cocrystals}) discuss the roughness evolution of 
	binary mixtures of rod--like compounds
	which are not phase--separating but may show the formation of co--crystals.
	In the second part (Sec.~\ref{sec:phaseseparation}) we will discuss the phase separating rod/sphere--shaped mixed films.

\section{Results}

\subsection{Roughness of intermixing thin films}
\label{sec:rough_intermixing}

In pure thin films, most of the studied compounds, i.e.\ pentacene (PEN), perfluoropentacene (PFP), diindenoperylene (DIP), sexithiophene (6T), and N,N'-bis(2-ethyl-hexyl)-1,7-dicyanoperylene-3,4/9,10-bis(dicarboxyimide) (PDIR-CN$_2$, where $R = C_8H_{17}$, branched) \cite{Belova_2017_JAmChemSoc} exhibit so-called layer-plus-island growth (Stranski-Krastanov, SK) on SiO$_2$ under the conditions employed here. In addition, picene (PIC) and fullerene (C60) exhibit typically island growth without wetting layer (Volmer-Weber, VW). The growth conditions of the single component films and the studied mixed films are summarized in supplementary Table 2. 

Fig.\,\ref{lab:fig:rough}a-f shows the roughness of six different types of molecular mixed films dependent on mixing ratio at a thickness of 20\,nm. All of the studied binary compounds mix on the molecular level, but they can be distinguished by their tendency to form a co-crystal \cite{Aufderheide_2012_PhysRevLett,Dieterle_2015_JPhysChemC,Dieterle_2017_PhysStatusSolidiRRL,Dieterle_2015_JPhysChemC,Reinhardt_2012_JPhysChemC,Broch_2013_JChemPhys,Hinderhofer_2011_JChemPhys}.  
The roughness of all pure materials is in general relatively large. A systematic trend in all films is that, upon mixing, the roughness is strongly decreased.

By comparing the roughness dependence of the various mixed systems, we can observe some significant differences. Mixtures with strong co-crystallization, i.e.\ the formation of an equimolar ordered co-crystal, exhibit a local roughness maximum at a 1:1 ratio (Fig.\,\ref{lab:fig:rough}e-f). In contrast, solid solutions do not show this local maximum (Fig.\,\ref{lab:fig:rough}a-b), which will be discussed in detail further below.

The strongest smoothing effect in absolute terms we find for mixtures with PIC (Fig.\,\ref{lab:fig:rough}b-c). Pure PIC exhibits strong island growth (VW) on SiO${_2}$ substrates \cite{Hosokai_2012_ChemPhysLett,Kurihara_2013_MolCrystLiqCryst,Hosokai_2015_JPhysChemC}. Upon mixing, the growth mode is apparently changed to SK mode. This effect is observable for PEN:PIC and PFP:PIC blends even for very low PEN or PFP concentrations.
The smoothing in mixtures is observed not only for mixtures with a compound showing strong islanding (SK : VW)  but also for mixtures where both compounds exhibit SK growth mode such as PEN, PFP and DIP. 
In order to better understand the overall picture, we discuss possible smoothing mechanisms and compare them to experimental results and theoretical simulations.

\subsection{Nucleation density}
\label{sec:nucleationdensity}
As a possible smoothing mechanism we consider first an increased nucleation density in the mixtures, which would yield a lower roughness via a simple geometric argument (supplementary Fig.\,1).\cite{Kotrla_2001_SurfSci}
The increase in nucleation density would lead in turn to a reduced in-plane correlation length $\xi$.
However, judging from the in-plane correlation length extracted from AFM data (supplementary Fig.\,1) we cannot identify a clear dependence between in-plane correlation length and roughness. Therefore we rule out this smoothing mechanism as being generally operational.

\begin{figure}[htb]
  \centerline{\includegraphics[width=7cm]{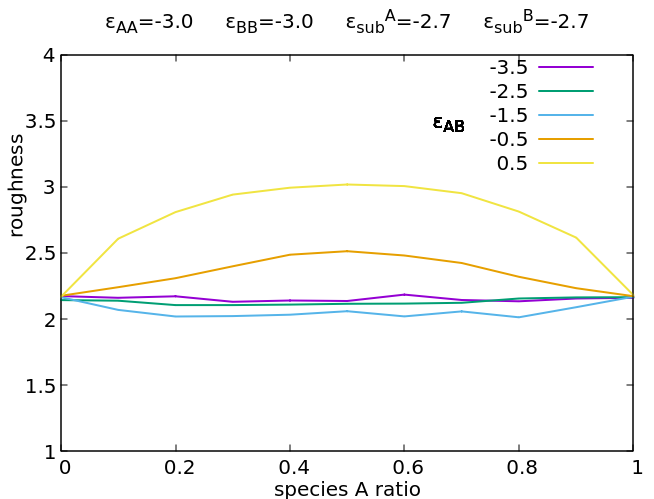}}
	\caption{Roughness in monolayers (ML) as a function of concentration of the species A after deposition of 15 ML. 
   Combination of two species with 3D growth, variation of $\epsilon_\text{AB}$. 
   Other parameters: $\Gamma=10^3$, $E^\text{ES}=3.0$, system size $300^2$, $E^\text{ES}_{ij}=E^\text{ES}$}
  \label{fig:eab_1}
\end{figure}

\subsection{Interaction strength and diffusion rate}
\label{sec:interactions_diffusion}
Two additional possible smoothing mechanisms may be that an increase in diffusion rate or a different interspecies interaction strength in the blend affects the roughness. 
On the one hand, a higher diffusion rate in the mixed film compared to the pure film would increase the hopping rate over step edges and lead potentially to smoother films. On the other hand, an increased interspecies interaction in the mixed films may reduce the roughness, because smoother films are energetically favored.

Experimentally, it is difficult to vary these two types of parameters systematically and independently.
This can be done in our generic model. To test the impact of interspecies interaction strength we set 
$D=D_\text{A}=D_\text{B}$ (Fig.~\ref{fig:eab_1}). We combine two species which show 3D growth (similar to PEN and DIP) and vary the interspecies energy from mixing conditions ($\epsilon_\text{AB}=-3.5$) to strong demixing ($\epsilon_\text{AB}=0.5$). Only in the demixing case, there is a noticeable increase of the roughness
(this is important for the rod--sphere mixtures discussed in Sec.~\ref{sec:phaseseparation}), 
otherwise it is insensitive. A simulation example for the combination of a material with 3D with an LBL growing material is shown in supplementary Fig.\,4.  

\begin{figure}[bt]
  \centerline{\includegraphics[width=7cm]{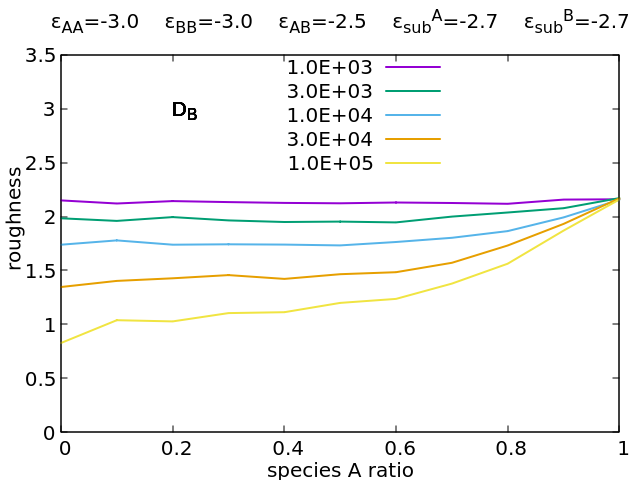}}
	\caption{Roughness in ML as a function of concentration of the species A after deposition of 15 ML. 
   Combination of one species with 3D growth with another which varies from 3D growth to LBL upon variation of $D_\text{B}$.
	The diffusion constant for the first species is set to $D_\text{A}/F=10^3$, corresponding to 3D growth. The diffusion constant for the 
second species is varied between $D_\text{B}/F=10^3$ (3D growth) and $10^5$ (LBL growth). 
   Other parameters: $E^\text{ES}=3.0$, system size $300^2$, $E^\text{ES}_{ij}=E^\text{ES}$.}
  \label{fig:db_1}
\end{figure}

To test the effect of the diffusion rate we choose energetic parameters which correspond to a well--mixing system (Fig.\,\ref{fig:db_1})
and increase the ratio of diffusion constants $D_\text{B}/D_\text{A}$.
The resulting roughness shows an approximately monotonic variation with concentration which, however, is not linear. Again, no decrease
of roughness upon addition of a second species is found as in the experiment. 
Nevertheless, $D_\text{B} \neq D_\text{A}$ implies that there could be different time scales for the development of 3D growth or the formation of 
islands. Thus one can expect fine--tuning effects as exemplified in the parameter sets for the picene mixtures where the
diffusion constant for picene was chosen such that it shows very strong island formation (supplementary Table 3).

We conclude that the roughness is insensitive to both diffusion rate and interaction strength (for mixing systems) in our growth regime  and cannot explain the drastic roughness decrease we observe experimentally. Therefore, another mechanism must play a role, which is addressed below.

\subsection{Step edge barrier}
\label{sec:stepedge}
Finally another possible reason for the reduced roughness might be a modified step edge barrier. 
For pure materials the step edge barrier is often significant and can lead to fast roughening \cite{Hlawacek_2013_JPhysCondensMatter,Yim_2007_JPhysChemC,Fendrich_2007_PhysRevB,Goose_2010_PhysRevB,Roscioni_2018_JPhysChemLett}. 

\begin{figure}[tbh]
  \centerline{\includegraphics[width=7cm]{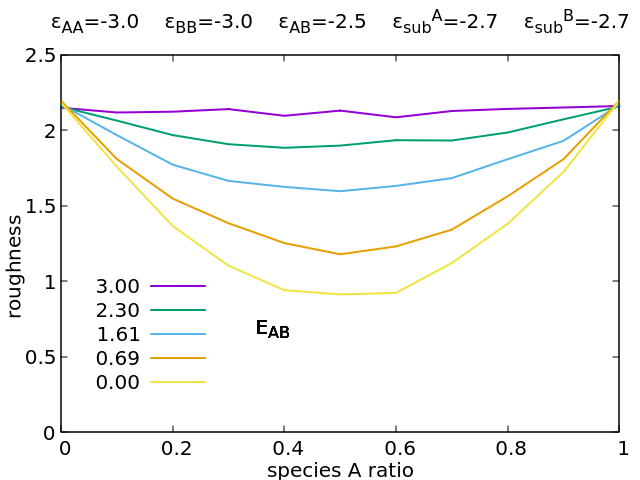}}
  \caption{Roughness in ML as a function of concentration of the species A after deposition of 15 ML. 
   Combination of two species with 3D growth. The ES barrier $E^\text{ES}_\text{AB}=E^\text{ES}_\text{BA}$ is varied from 0.0 to 3.0, i.e. interlayer hops of one species on top (or from) a particle from the other species are more likely.
   Other parameters: $\Gamma=10^3$, $D=D_\text{A}=D_\text{B}$, $E^\text{ES}_{ii}=3.0$, system size $300^2$.}
  \label{fig:es_1}
\end{figure}

The variation of the interspecies step edge barrier ($E^\text{ES}_{ij}$) in the simulations leads to the generic roughness effect seen in the experiment, i.e. the reduction of roughness upon mixing (Fig.~\ref{fig:es_1}). We have seen that for energetic conditions suitable for mixing there is no strong variation of the roughness with the interspecies energy. In this case, the condition $E^\text{ES}_{ij} < E^\text{ES}_{ii}$ is the only possible cause of roughness reduction. We conclude that for a reduced and species dependent step edge barrier our KMC simulations are in excellent agreement with experimental data for the rod-rod mixed systems without co-crystallization (Fig.\,\ref{lab:fig:rough}a-d).

\begin{figure*} [bth]
	\begin{center}
	\includegraphics [width=14cm] {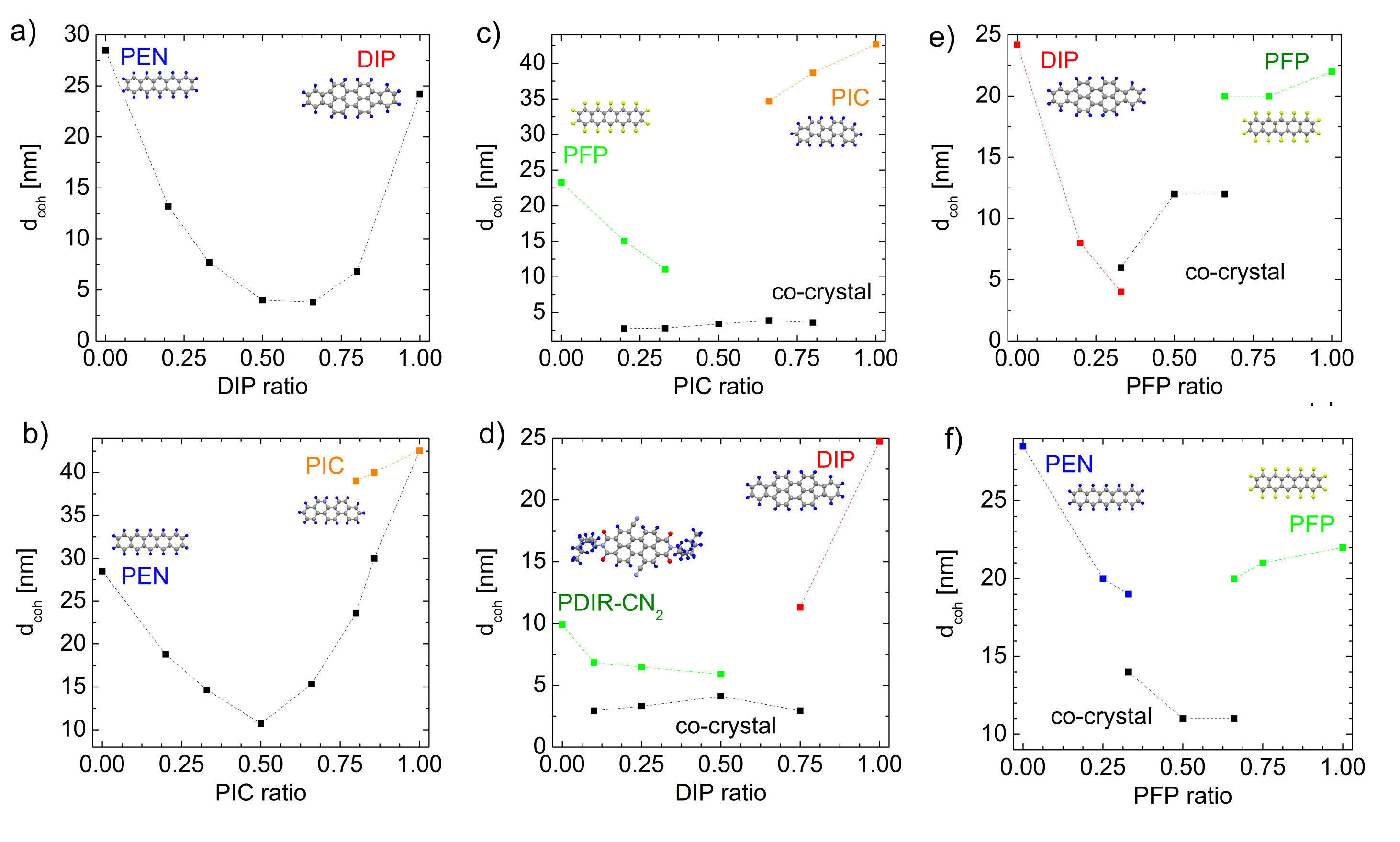}
	\caption{Coherent island size $d_{coh}$ with rod like compounds dependent on mixing ratio: a) PEN:DIP\cite{Aufderheide_2012_PhysRevLett} b) PEN:PIC\cite{Dieterle_2015_JPhysChemC,Dieterle_2017_PhysStatusSolidiRRL} c) PFP:PIC\cite{Dieterle_2015_JPhysChemC} d) DIP:PDIR\cite{Belova_2017_JAmChemSoc} e) DIP:PFP\cite{Reinhardt_2012_JPhysChemC,Broch_2013_JChemPhys} f) PEN:PFP\cite{Hinderhofer_2011_JChemPhys}. From a)-f) the in-plane co-crystallization is increasing, i.e.\ PEN:DIP (a) are nearly randomly intermixing, whereas PEN:PFP (f) exhibits well ordered co-crystallization and b)-e) are intermediate cases.}
	\label{lab:fig:rough2}
	\end{center}
\end{figure*}

To rationalize the lowered step edge in the blends, we recall that the step edge barrier should be viewed as an effective quantity
that arises through a weighted average over different step conformations \cite{TMichely_2004_book}. In particular, molecular thin films have a distribution of 
different step edge barriers, dependent on the crystal orientation and the trajectory of the diffusing molecule over the step edge \cite{Hlawacek_2013_JPhysCondensMatter}.
 
For a well-defined pure crystalline domain the step edge barrier is relatively high.
When we randomly introduce guest molecules into a molecular crystal (and also at the domain boundaries), the 
number of possible step conformations will increase, since guest molecules do not fit exactly into the lattice of the host and 
distort it. Therefore, it is natural to assume that with an increasing amount of guest molecules the distribution of step edge barriers
broadens and some barriers are lower compared to the pure crystal. 
This effect should be strongest for large mixing ratios ($\approx$ 1:1). Since the diffusion lengths of organic molecules are relatively large \cite{Heringdorf_2001_NatureLondonUK}, the introduction of only a few low potential barriers, has a strong impact on the molecular downhill transport. Thus, the film roughening will be reduced.

It should be noted that this scenario is distinct from the well-known effect of surfactant molecules in metal homoepitaxy, which segregate at step edges and systematically modify the barrier for descending atoms\cite{Esch_1994_PhysRevLett}. Here, we do not expect the step edge barriers in the mixed films to be lower on average. Rather,
the broadening of the distribution of barriers induced by the molecular disorder opens pathways for facile descent that are preferentially used by thermal activation. Since the solid-on-solid model does not account for the orientational degrees of freedom of the molecules, in our simulations this effect is nevertheless represented by an overall reduction of the barrier.

As a measure for the increased defect density we use the coherently scattering island size $d_{coh}$ of the blends (Fig.\,\ref{lab:fig:rough2}) which is derived from the FWHM of in-plane Bragg reflections. We observe that indeed $d_{coh}$ decreases in the blends correlated to the decreasing roughness. For example, for statistically intermixing compounds such as PEN:DIP (Fig.\,\ref{lab:fig:rough}a), we find a minimum of $d_{coh}$ near the 1:1 ratio (Fig.\,\ref{lab:fig:rough2}a), which is consistent with the above explanation of the generation of smaller step edge barriers by guest molecules.

\subsection{Impact of co--crystal formation}
\label{sec:cocrystals}
The observation of a local maximum in the roughness can again be discussed in terms of the in-plane crystallinity (Fig.\,\ref{lab:fig:rough2}).
For PEN:PFP and DIP:PFP, we find a strong tendency towards the formation of a 1:1 co-crystal with a relatively large $d_{coh}>10$\,nm. Excess molecules of either compound phase separate in pure domains \cite{Hinderhofer_2011_JChemPhys,Breuer_2013_JChemPhys,Reinhardt_2012_JPhysChemC,Broch_2013_JChemPhys}.
Due to this growth behavior, the crystallinity is increased at 1:1 ratio in these mixtures in comparison to statistically mixed compounds. Then, the effect of low potential step edge barriers introduced by  guest molecules is weaker. The observation of increased roughness with higher crystallinity, supports the assumption that low step edge barriers are the main smoothing mechanism in organic mixed films of two rod-like molecules.

For the simulations strong mixing conditions are characterized by $\epsilon_\text{AB} \ll (\epsilon_\text{AA}+\epsilon_\text{BB})/2$. In that limit the lattice model shows a stable checkerboard phase which is similar to the 1:1 co--crystal formed in the PFP mixtures. 
There is one important difference. Experimental PFP mixtures not at equal (1:1) concentrations show phase separation into
a pure component and the 1:1 co--crystal. The lattice model does not show a similar phase separation, rather, the checkerboard
structure is randomly mixed in the system. 

\begin{figure}[bt]
  \centerline{\includegraphics[width=7cm]{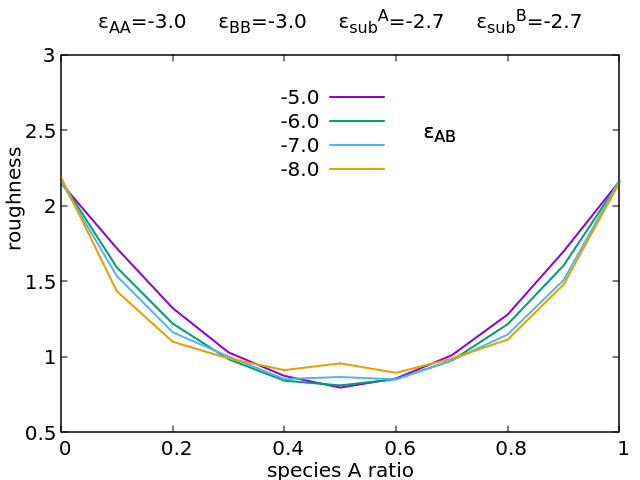}}
	\caption{Roughness in ML as a function of concentration of the species A after deposition of 15 ML. 
   Combination of two species with 3D growth, variation of $\epsilon_\text{AB}$ in the strong mixing regime. 
   Other parameters: $\Gamma=10^3$, $E^\text{ES}_{ii}=3.0$, $E^\text{ES}_{ij}=0.0$ ($(i\neq j)$), system size $300^2$.}
  \label{fig:eab_4}
\end{figure}

Since the co--crystallization is the most important difference of the PFP mixtures with PEN and DIP compared to the other
mixing blends with PEN and/or DIP, we take the PEN:DIP parameters (including unequal ES barriers) but decrease $\epsilon_\text{AB}$ substantially.
The result is shown in Fig.\,\ref{fig:eab_4}. Overall, there is no substantial change to the PEN:DIP curve but curiously for the
lowest $\epsilon_\text{AB}$ a small hump is forming for $c_\text{A}=0.5$. The effect seems to be genuine and persists also for a choice of less unequal ES barriers, nevertheless it is too small compared to the experimentally seen effect.

Therefore, the PFP:DIP and PFP:PEN mixtures should rather be considered as weakly phase separating mixtures of the pure compound with the respective  1:1 co-crystal, and the roughness behavior of these mixtures is more similar to those of weakly phase separating mixtures like DIP:C$_{60}$ described below.

\subsection{Sphere-like with rod-like compound: Phase separation}
\label{sec:phaseseparation}
Qualitatively different from intermixing rod/rod blends discussed above are rod/sphere such as mixtures realized with C60 and a rod-shaped compound. Due to geometrical constraints rod/sphere blends are typically phase separating in thermal equilibrium, but due to kinetic effects partially intermixed in thin films \cite{Banerjee_2013_PhysRevLett}.   
Fig.\,\ref{lab:fig:rough}g-i) shows the roughness of mixed films of C60 with three different rod-shaped compounds (DIP, 6T, PEN). 
We observe that for a small amount of guest molecules the roughness is decreased, which can be explained consistently by a reduced in-plane crystallinity. 

For a phase separating system, the roughness depends on the domain size and therefore the degree of phase separation, which is related to the interspecies interaction energy \cite{Kotrla_1998_PhysRevB}.
From KMC simulations we conclude, that the smoothing effect induced by a low step edge barrier (Fig.\,\ref{fig:es_1}) and the roughening effect induced by unfavorable interspecies energies (Fig.\,\ref{fig:eab_1}, supplementary Fig.\,4), i.e.\ phase separation, are competing effects. Both effects can be active to a different degree dependent on mixing ratio.

The three mixtures studied (Fig.\,\ref{lab:fig:rough}g-i) exhibit nano phase separation into pure domains dependent on the growth conditions \cite{Salzmann_2008_JApplPhys,Banerjee_2013_PhysRevLett,Lorch_2015_ApplPhysLett,Lorch_2016_JApplCrystallogr}. 
For DIP:C60 and 6T:C60 and mixing ratios deviating from 1:1, we observe smaller roughness values compared to the pure compounds, induced due to larger disorder and therefore smaller step edges as described above. For DIP:C60 the driving force for phase separation (dependent on the interspecies energy) is weak, resulting in small $d_{coh}$ (supplementary Fig.\,2).
In contrast, for PEN:C60, apparently the interspecies energies are strongly unfavorable for mixing, leading to a large $d_{coh}$ and the largest roughness of the three systems studied.

These observations are also consistent with KMC simulations with a variation of the ES barrier (Fig.\,\ref{fig:es_1}). For energetic conditions suitable for demixing we have seen that the roughness of blended films is higher, and the degree of roughness increase depends on the propensity for phase separation (i.e. the value of $\epsilon_\text{AB} - (\epsilon_\text{AA}+\epsilon_\text{BB})/2$) but also on the single species growth mode. Therefore we have here two competing mechanisms influencing the final roughness. For weakly phase separating systems the ES effect could dominate but for strongly phase separating systems it can be the other way around. This is also seen in the experiments.

\begin{figure*} [bt]
	\begin{center}
	\includegraphics [width=17cm] {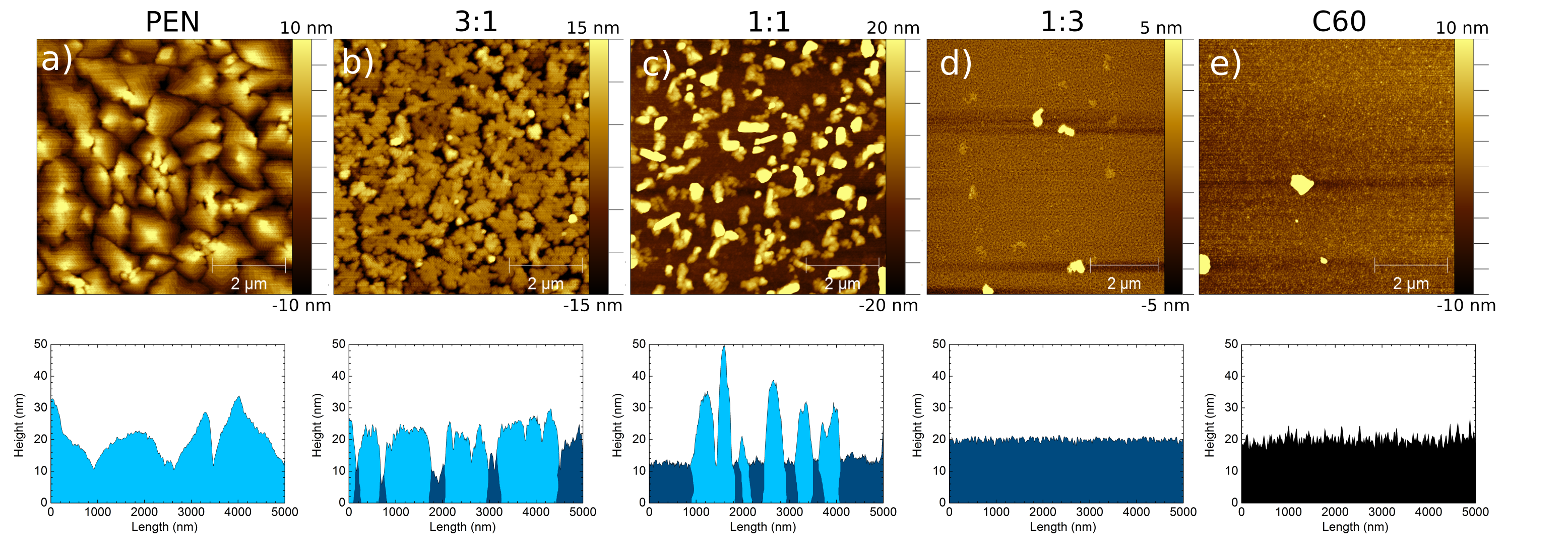}
	\caption{AFM images of a) PEN, e) C$_{60}$ and three PEN:C$_{60}$ mixtures b-d) with different mixing ratios. Sketches below each AFM image show typical line scans. Colors illustrate the domain compositions: light blue (pure PEN), dark blue (nano phase separated mixture), black (C60).}
	\label{lab:fig:AFM}
	\end{center}
\end{figure*}

For 1:1 mixing ratios we find a local roughness maximum for all three mixed systems. For DIP:C60 it was shown that at 1:1 mixing ratios the films exhibit two types of domains: a nano phase separated wetting layer and pure domains of DIP \cite{Banerjee_2018_JPhysChemC}. Similarly, in AFM data of PEN:C60 (Fig.\,\ref{lab:fig:AFM}) we also observe a pronounced 3D growth of pure PEN domains near the 1:1 mixing ratio in combination with a mixed wetting layer \cite{Salzmann_2008_JApplPhys}. The strong 3D growth of the pure domains in these films are the main cause for the local roughness increase. Since the lateral separation between these domains is on the order of $\approx 1000$ lattice sites, this effect cannot be completely captured by our KMC simulations. 
However, we find that the increase in roughness at the 1:1 ratio scales overall with the driving force for phase separation (weak for DIP:C60, strong for PEN:C60) and depends presumably mainly on the interspecies interaction energies.
We conclude that the roughening and smoothing mechanisms are qualitatively the same for all three mixtures studied, but quantitatively of course dependent on material properties.

\section{Discussion}
We presented an extensive and systematic study on the roughness evolution of different organic mixed thin films. We distinguished the roughening mechanism in intermixing rod/rod blends and phase separating rod/sphere blends. KMC simulations revealed that a species dependent step edge barrier is the main smoothing mechanism. 
As a possible scenario we propose a broadening of the distribution of step edge barriers induced by guest molecules. This idea is supported by the strong correlation between in-plane coherent crystal length and roughness for all studied blends.
For intermixing rod/rod blends we find a roughness minimum close to the 1:1 ratio induced by a reduced step edge barrier. For rod/sphere blends the roughness depends, in addition to the reduced step edge barrier, on the competing effect of phase separation. Finally, a local roughness maximum at 1:1 ratio was found also for co-crystallizing blends. These blends behave similar to weakly phase separating blends, where phase separation occurs between a 1:1 co-crystal and the pure compound. 

Our study shows an intriguing and subtle connection between non-equilibrium structure formation and equilibrium phase behavior
mediated by the kinetics of interlayer transport. Importantly, the near-universal smoothing observed in the mixed films
relies on the thermally activated character of step crossing events, which implies that the transport is effectively dominated by 
the lowest available barriers. We expect that similar scenarios may be found in other systems where complex molecular interactions
give rise to a broad distribution of kinetic rates.

\section{Methods}
\subsection{Thin film preparation}
All films studied have a thickness of 20\,nm and were prepared by thermal evaporation in vacuum onto Si wafers with native oxide layer. 
Thin films were deposited on silicon wafers with native SiO$_2$ (surface roughness $\sigma_{\mathrm{rms}} = 0.3$\,nm) under ultra high vacuum (UHV) conditions (base pressure $<1 \cdot 10^{-7}$\,Pa) by thermal evaporation.\cite{Ritley_2001_RevSciInstrum}
Before deposition, substrates were cleaned ultrasonically with acetone, isopropyl alcohol, and ultra pure water, followed by heating to 700\,K in the UHV growth chamber. All films were deposited at a substrate temperature of $T \sim 300$\,K. The growth rate was monitored by a quartz crystal microbalance. Typical evaporations rates are shown in supplementary Table 1.

\subsection{X-ray Scattering}
X-ray Reflectivity (XRR) and grazing incidence X-ray diffraction (GIXD) were measured either at the X04SA beamline of the Swiss Light Source, Paul Scherrer Institut, Villigen, Switzerland or at beamline ID10 of the ESRF in Grenoble, France.

\subsection{Roughness Determination}
Roughness ($\sigma$) values were determined by fitting XRR with Motofit\cite{Nelson_2006_JApplCrystallogr} and from AFM with Gwyddion\cite{Nevcas__}. Both methods yielded very similar results. We estimate the error bars for the $\sigma$ values on the range of 10\%. 

\subsection{In-plane coherent crystal size}
Lower limits of the in-plane coherent crystal sizes $d_{coh}$ were determined by the Scherrer formula $l_{s} = 2 \pi \cdot (\mathrm{FWHM})^{-1}$, where  $\mathrm{FWHM}$ is the full width half maximum of the peak in \AA$^{-1}$ determined with a Gaussian fit-function.\cite{Smilgies_2009_JApplCrystallogr} The instrumental broadening of the diffractometer was not included in the calculation, therefore only lower limits of $l_{s}$ are given.

\subsection{In-plane correlation length}
In-plane correlation lengths were determined with Gwyddion\cite{Nevcas__} from AFM data by fitting the one-dimensional power spectral density function (PSDF) with a power law. The used PSDF for each sample is an average from 2-4 images with sizes between $3-10 \mu$m. 

\begin{figure}
	\begin{center}
		\includegraphics[width=8cm]{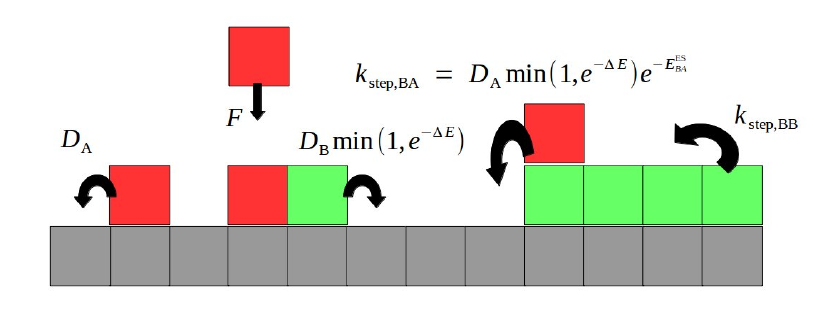}
		\caption{Hopping and insertion moves with associated rates in the KMC model.}
		\label{fig:moves}
	\end{center}
\end{figure}

\subsection{Simulation Model}
\label{sec:sim_model}
For the KMC simulations, we employ a simple film growth model using a binary lattice gas (species A and B) on a cubic lattice 
with interaction energy parameters as illustrated in Fig.~\ref{fig:energetics}.

Deposition on top of the film or the bare substrate at random substrate plane coordinates is controlled by a rate $F$
(particles per unit time and lattice site). Diffusion respects the solid--on--solid condition: only the
particles (species $i$) in the top layer are allowed to diffuse to a lateral next--neighbor site with rate 
$D_i \;\text{min}(1,\exp( -\Delta E))$ where $D_i$ is a species--dependent free diffusion constant and 
$\Delta E$ is the energy difference between final and initial state in units of $k_B T$. 
In such a move, particles of species $i$ may also ascend (moving on top of a particle of species $j$) or descend one layer 
(moving down from a particle of species $j$) in which case the rate is multiplied with 
$\exp(-E^\text{ES}_{ij})$ where $E^\text{ES}_{ij}$ is an Ehrlich--Schwoebel (ES) barrier. Neither overhangs nor desorption are allowed.
The moves are illustrated in Fig.\,\ref{fig:moves}.

In the one--component case (supplementary Fig.\,3), the model is characterized by the four constants $\epsilon=\epsilon_\text{AA}, \epsilon_\text{sub}=\epsilon_\text{sub}^\text{A}, E^\text{ES}=E^\text{ES}_\text{AA},
\Gamma = D_\text{A}/F$. Actual growth experiments of organic thin films are characterized by $|\epsilon|= 10 \dots 15$,
and $\Gamma = 10^9 \dots 10^{11}$ which is difficult to simulate, owing to computational costs. 
However, at lower energies and smaller ratios
$\Gamma$  the model shows similar growth modes as seen experimentally. These are (a) island growth from the start (ISL) when
$\epsilon_\text{sub}$ is low enough, (b) layer--by--layer growth (LBL) and (c) 3D growth of varying degree. 
The model shows two characteristic transitions
which, however, are not sharp:
(i) ISL-LBL which for given $\epsilon,E^\text{ES}$ depends on both $\epsilon_\text{sub}, \Gamma$ and whose order parameter can
be characterized by the coverage difference of layer 1 and 2 after depositing only 1 ML. 
(ii) LBL-3D resp. ISL-3D which for given $\epsilon,\epsilon_\text{sub},E^\text{ES}$
depends on $\Gamma$ and where a suitable order parameter is e.g. the integral of the anti--Bragg intensity and 
needs deposition of tens of MLs for locating.

More details on the one--component growth modes and the associated transitions can be found in Ref.~\cite{empting20}.

\subsection{Simulation Parameter Selection}
In selecting the parameters for the pure systems, we proceed with the assumption that there are scaling relations
for the temporal roughness evolution, i.e. that simulations at lower $|\epsilon|, |\epsilon_\text{sub}|, E^\text{ES}, \Gamma$ correspond also to certain sets of these parameters with higher values. 
In the literature, the epitaxial case $\epsilon=\epsilon_\text{sub}$ and $E^\text{ES}=0$ has been investigated recently \cite{Assis_2015_JStatMech} and
a scaling $r \propto L^\beta/(\Gamma^{3/2}(\exp(-|\epsilon|) +a)$ has been found ($r$ is the normalized roughness with for a layer thickness of 1, $L$ is the number of deposited layers, $\beta \approx 0.2$, $a = 0.025$).

We have investigated the scaling relations for the exemplary case of the PEN:DIP mixtures for energy parameters $\epsilon = {-3} \dots {-5}$ and a wider range of diffusion parameters $\Gamma = D_A = D_B = 10^3\dots10^6$. 
We found that the composition dependence of a multilayer film depends rather well on the single variable $\Gamma^{1.5} \exp(-|\epsilon|)$ which was also found in the single-component case by Assis. Thus, e.g., when coming from a more  realistic energy scale of  $\epsilon=-15$ to  $\epsilon=-3$, one may reduce $\Gamma$ by a factor of $2 \times 10^3$ if that scaling holds. 

In going to growth of binary systems, the dimension of the parameter space of this simple model is already enlarged to 10 
(4 parameters for each of the pure systems, the cross--species energy $\epsilon_\text{AB}$ (which controls mixing and demixing) and
the cross--Ehrlich--Schwoebel barrier $E^\text{ES}_\text{AB}=E^\text{ES}_\text{BA}$). 
We concentrated on combining different pairs of one--component growth modes which reflect the experimental material 
combinations. In general we found that for the simple choices
$\epsilon_\text{AB} \approx (\epsilon_\text{AA}+\epsilon_\text{BB})/2$, $D_\text{A} \approx D_\text{B}$ and $E^\text{ES}_{ij}=\text{const}$ the roughness properties of the films
linearly interpolate between those of the pure substances. Therefore, any more prominent mixture effects can only be expected 
when deviating from these choices.

Simulations were done on a grid size of $M=300$. Our tests for larger grid sizes ($M=800$) show that the obtained roughness values do not depend strongly on the grid size (supplementary Fig.\,5).

\section{Acknowledgment}
Support from the DFG and the BMBF is gratefully acknowledged. We thank M. Kotrla for fruitful discussions and numerous members of the T\"ubingen group
for contributing data.

\bibliography{bibfile2}				

\end{document}



\author{Alexander Hinderhofer} \affiliation{Institute for Applied Physics, University of T\"ubingen, Auf der Morgenstelle 10, 72076 T\"ubingen, Germany}
\author{Jan Hagenlocher} \affiliation{Institute for Applied Physics, University of T\"ubingen, Auf der Morgenstelle 10, 72076 T\"ubingen, Germany}
\author{Alexander Gerlach} \affiliation{Institute for Applied Physics, University of T\"ubingen, Auf der Morgenstelle 10, 72076 T\"ubingen, Germany}
\author{Joachim Krug} \affiliation{Institute for Biological Physics, University of Cologne, Z\"ulpicher Strasse 77, 50937 K\"oln, Germany}
\author{Martin Oettel} \affiliation{Institute for Applied Physics, University of T\"ubingen, Auf der Morgenstelle 10, 72076 T\"ubingen, Germany}
\author{Frank Schreiber} \affiliation{Institute for Applied Physics, University of T\"ubingen, Auf der Morgenstelle 10, 72076 T\"ubingen, Germany}

\date{\today}
\title{Non-equilibrium Roughness Evolution of Small Molecule Mixed Films Reflecting Equilibrium Phase Behavior}

\maketitle

\section{Supplementary Tables}

\begin{table}[htb]
\caption{Overview of growth rates of the studied mixed films.}
\begin{tabular} {cc}
\hline
Mixture		 		&  growth rate (nm/min)		  		\\ \hline
PEN:DIP				&  0.2	 \\ 
PEN:PIC				&  0.3	 \\
PFP:PIC				&  0.3	 \\
PDIR:DIP			&  0.5	 \\
DIP:PFP				&  0.1	 \\
PEN:PFP				&  0.2   \\ 
DIP:C60				&  0.4	 \\
6T:C60				&  0.4	 \\
PEN:C60				&  0.2	 \\
\hline
\end{tabular}
\label{tab0}
\end{table}

\begin{table}[htb]
\caption{Overview of studied mixed film systems categorized by their growth mode (Stranski-Krastanov or Volmer-Weber) in the pure compounds and the equilibrium phase behavior of the mixture.}
\begin{tabular} {lccc}
\hline
							&  co-crystal		  	& solid solution  	& phase separation			\\ \hline
SK : SK				&  DIP:PFP			  	& PEN:DIP   				&  \\ 
							&  PEN:PFP					& 									& \\
							&  DIP:PDIR-CN$_2$	& 									& \\
							&  									& 									& \\
VW : SK				&  PIC:PFP  				& PIC:PEN   				& C60:PEN    			\\ 
							&  									& 									& C60:DIP					\\
							&  									& 									& C60:6T					\\
\hline
\end{tabular}
\label{tab1}
\end{table}

\begin{table}[h]
\begin{tabular*}{\textwidth}{c @{\extracolsep{\fill}} llllllllllllcc}
 \hline \hline \\
 mixture & Fig. & $\epsilon_\text{AA}$ & $\epsilon_\text{BB}$ & $\epsilon_\text{AB}$ & $\epsilon_\text{sub}^\text{A}$ & $\epsilon_\text{sub}^\text{B}$ & $D_\text{A}$  & $D_\text{B}$ & $E^\text{ES}_\text{AA}$ & $E^\text{ES}_\text{BB}$ & $E^\text{ES}_\text{AB}$ &\multicolumn{2}{c}{single spec.} \\
        &       &        &         &       &          &         &         &       &             &             &            & \multicolumn{2}{c}{growth mode}  \\   
        &       &        &         &       &          &         &         &       &             &             &            &  1 & 2  \\ \\ \hline  
PEN:DIP & 1a    & -3.0   & -3.0   & -2.5  &  -2.7     & -2.7     & $10^3$ & $10^3$ & 3.0        & 3.0         & 0.0  & 3D & 3D \\  
PIC:PEN & 1b    & -2.0   & -2.0   & -2.5  &  -1.0     & -1.9     & $10^5$ & $10^3$ & 0.69       & 3.0         & 0.0  & isl & 3D\\  
PIC:PFP & 1c    & -2.0   & -2.0   & -1.5  &  -1.0     & -1.9     & $10^5$ & $10^3$ & 0.69       & 3.0         & 0.0  & isl & 3D \\  
DIP:PDIR& 1d    & -3.0   & -3.0   & -2.5  &  -2.7     & -2.7     & $10^3$ & $10^3$ & 3.51       & 1.39         & 0.0 & 3D & 3D (weak) \\  
PFP:DIP & 1e    & -3.0   & -3.0   & -8.0  & -2.7     & -2.7     & $10^3$ & $10^3$ & 3.0        & 3.0          & 0.0  & 3D & 3D \\  
PFP:PEN & 1f    & -3.0   & -3.0   & -8.0  & -2.7     & -2.7     & $10^3$ & $10^3$ & 3.0        & 3.0          & 0.0  & 3D & 3D \\  
DIP:C60 & 1g    & -3.0   & -3.0   & -1.5  & -2.7     & -1.0      & $10^4$ & $10^4$ & 3.0        & 3.0          & 0.0 & 3D & isl \\  
6T:C60  & 1h    & -3.0   & -3.0   & -0.5  & -2.7     & -1.0      & $10^4$ & $10^4$ & 3.0        & 3.0          & 0.0 & 3D & isl \\  
PEN:C60  & 1i    & -3.0   & -3.0   & +0.5  & -2.7     & -1.0      & $10^4$ & $10^4$ & 3.0        & 3.0          & 0.0  & 3D & isl\\ \hline \hline
\end{tabular*}
	\caption{Simulation parameters used for the results presented in Fig. {2} in the main paper. The growth modes for the single species have been 
read off from the roughness behavior after growth of 15 ML.``3D'' refers to 3D growth which is characterized by a monotonic increase of roughness
with deposited ML. ``3D(weak)'' is also 3D growth with a weaker increase of the roughness. ``isl'' refers to island growth which in the
model is characterized by an initially steeply increasing roughness. It may then smoothly cross over to 3D growth, but it may also decrease again
at higher deposition when islands begin to merge.}
\label{tab:parameters}
\end{table}

\clearpage

\section{Supplementary Figures}

\begin{figure*} [hbt]
	\begin{center}
	\includegraphics [width=5cm] {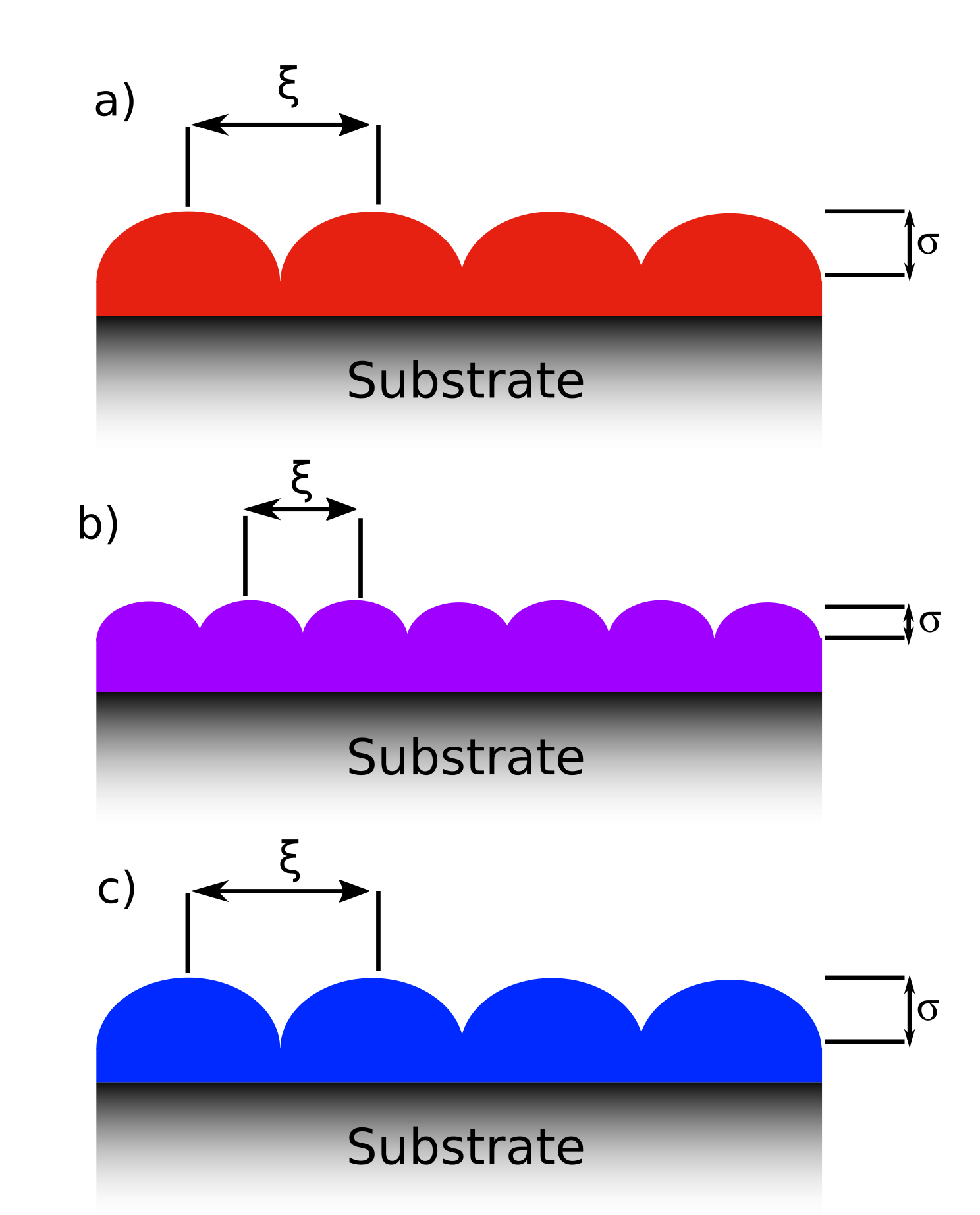}
	\includegraphics [width=11cm] {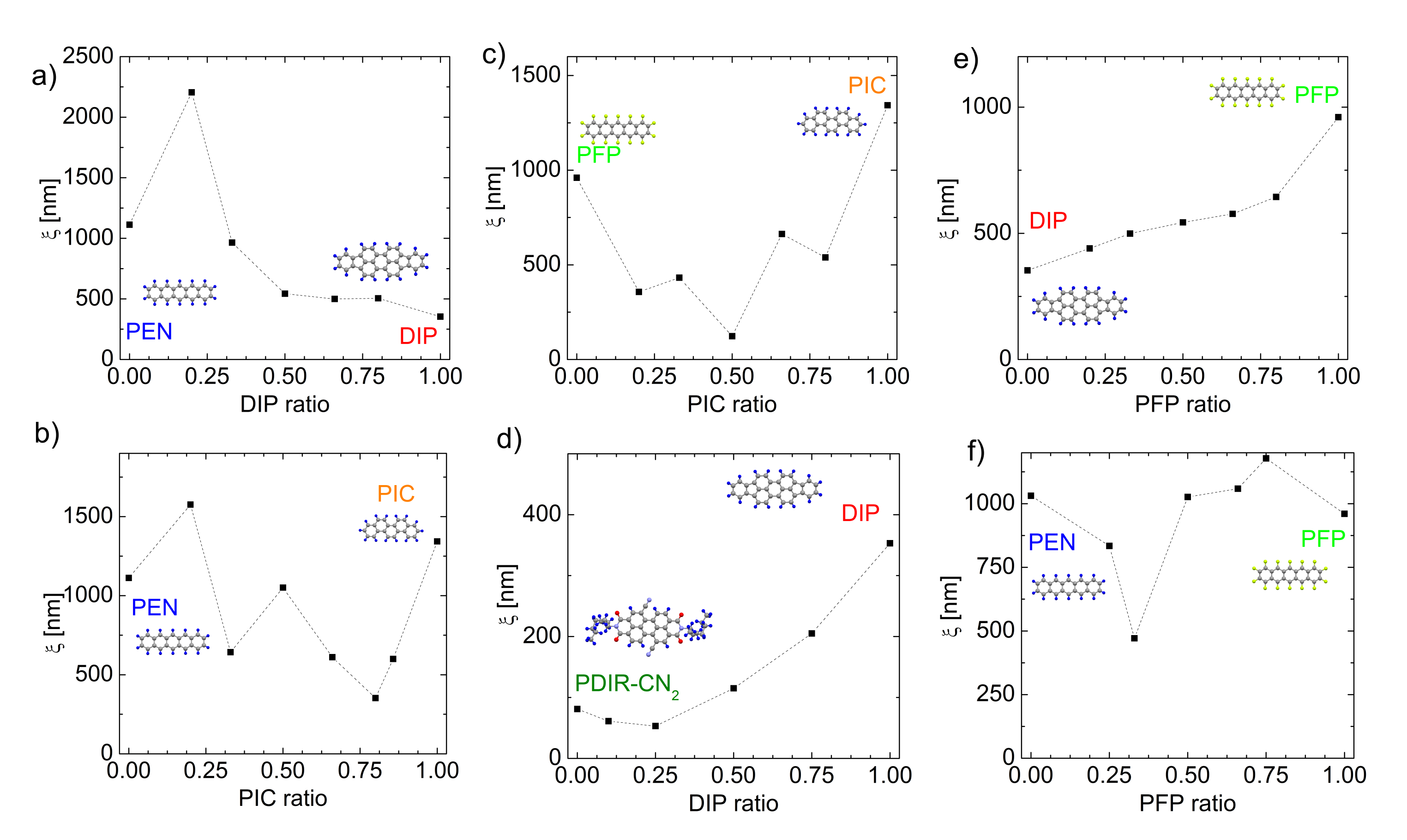}
	\caption{\textbf{Left:} Sketch of roughness reduction by increase of nucleation density. a) and c) pure materials, b) 1:1 mixture has reduced in-plane correlation length $\xi$. \textbf{Right:} Correlation length of blends with rod like compounds dependent on mixing ratio extracted from AFM images. a) PEN:DIP b) PEN:PIC c) PFP:PIC d) DIP:PDIR e) DIP:PFP f) PEN:PFP. It is obvious that for some mixed systems (e.g. PFP:PIC) there appears to be a nearly linear correlation between $\xi$ and $\sigma$, however for other mixtures (PEN:PIC) both parameters seem to be uncorrelated. Especially for PEN:DIP, the correlation length in 4:1 mixture is very large, however the roughness is very low. Judging from these observations we conclude that the island density may play a role for the changed roughness for some material systems but cannot explain the general roughness behavior described in the main text.}
	\label{lab:fig:corr}
	\end{center}
\end{figure*}

\clearpage

\begin{figure} [ht]
	\begin{center}
	\includegraphics [width=8cm] {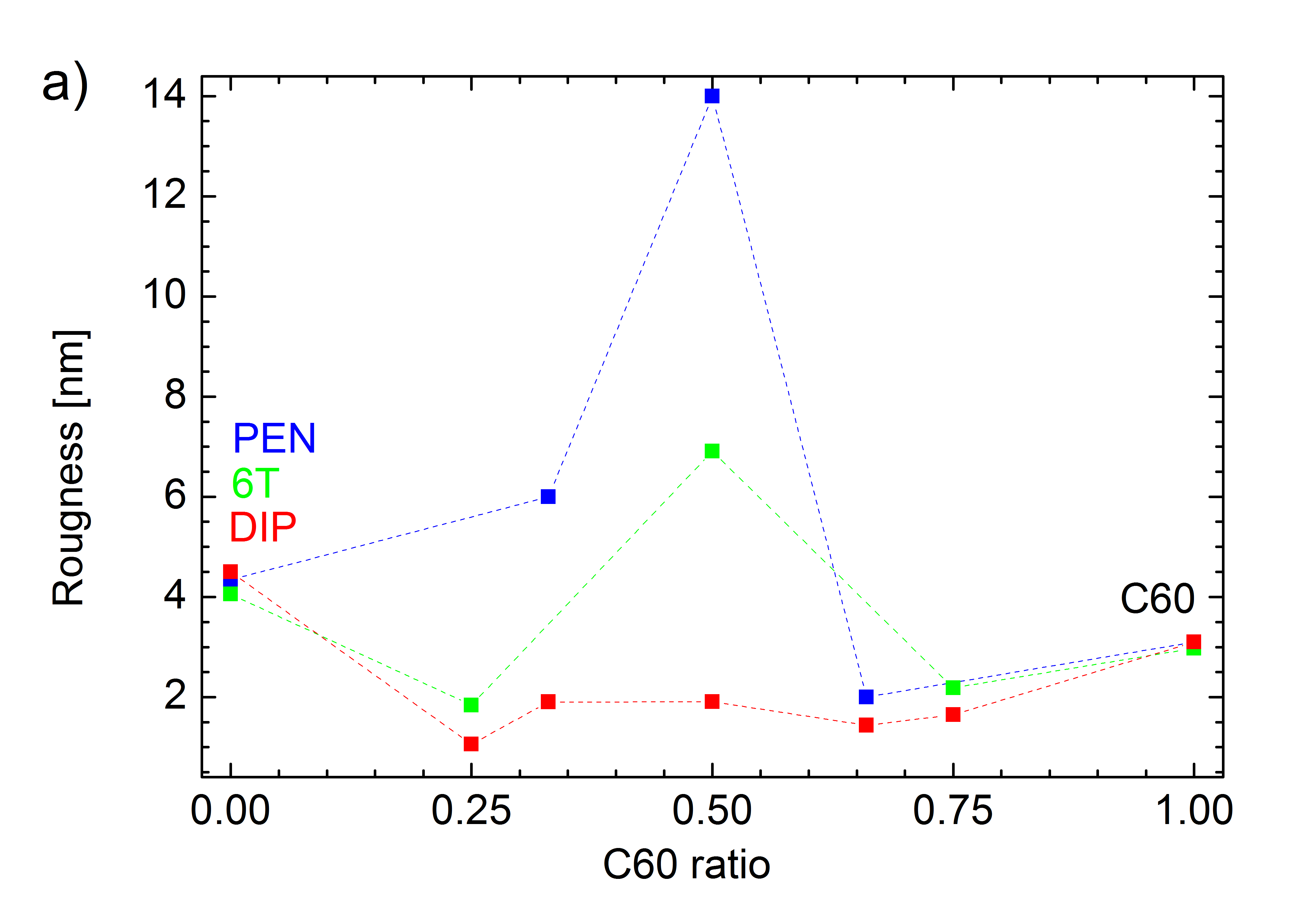}
	\includegraphics [width=8cm] {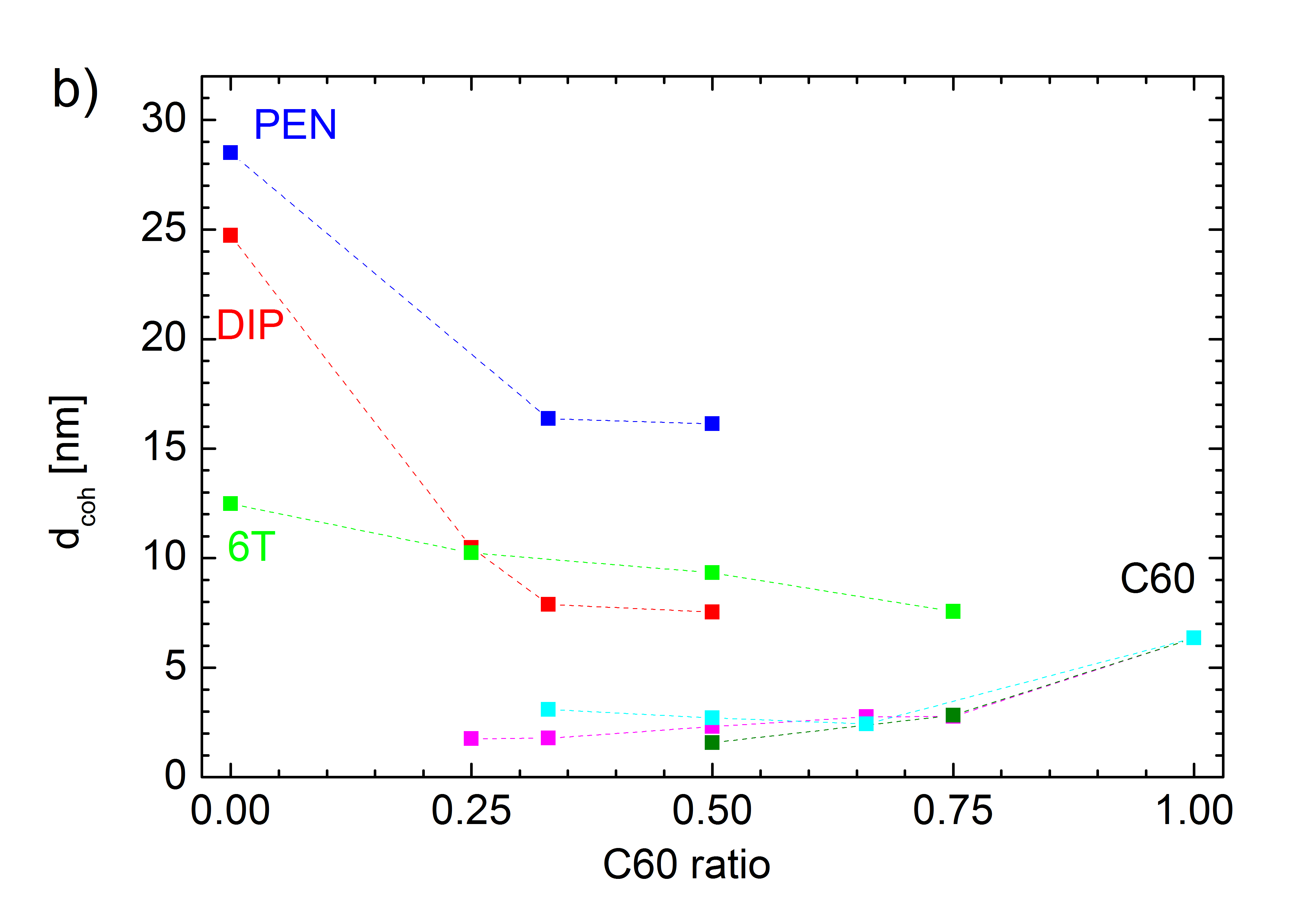}
	\caption{a) Roughness $\sigma$ of blends of C$_{60}$ with a rod like compound. b) In-plane coherent island size $d_{coh}$ dependent on the mixing ratio of rod/sphere mixtures determined from GIXD.}
	\label{lab:fig:C60DIP}
	\end{center}
\end{figure}

\clearpage

\begin{figure}[hb]
  \centerline{\includegraphics[width=8cm]{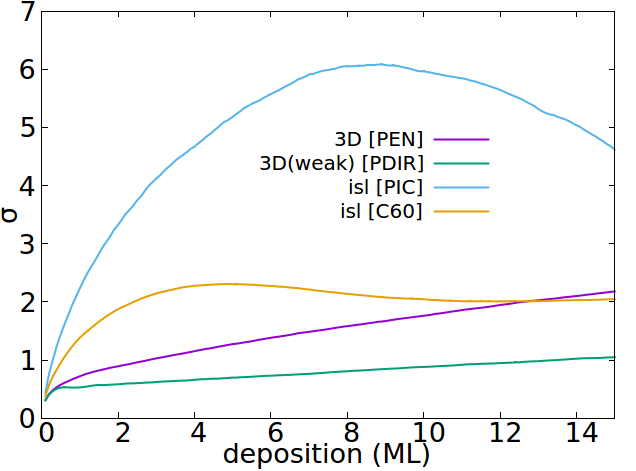}}
  \caption{
Roughness $\sigma$ as a function of deposition (in ML) for 4 typical pure components as PIC, PDIR, PIC and C60 described in the model, system size $300^2$.}
  \label{fig:singlespec}
\end{figure}

\clearpage

\begin{figure}[ht]
  \centerline{\includegraphics[width=8cm]{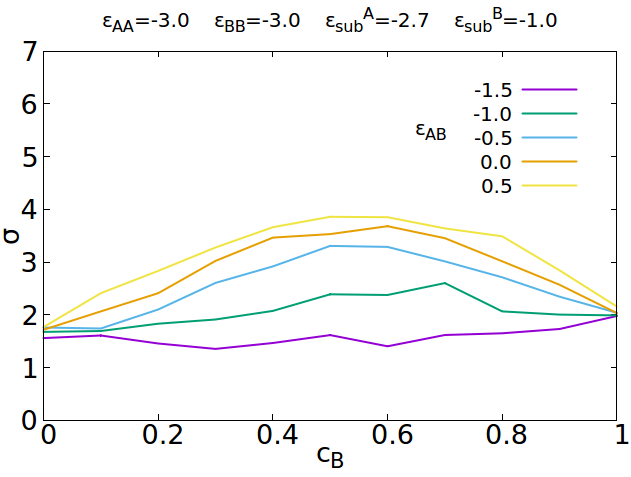}
              \includegraphics[width=8cm]{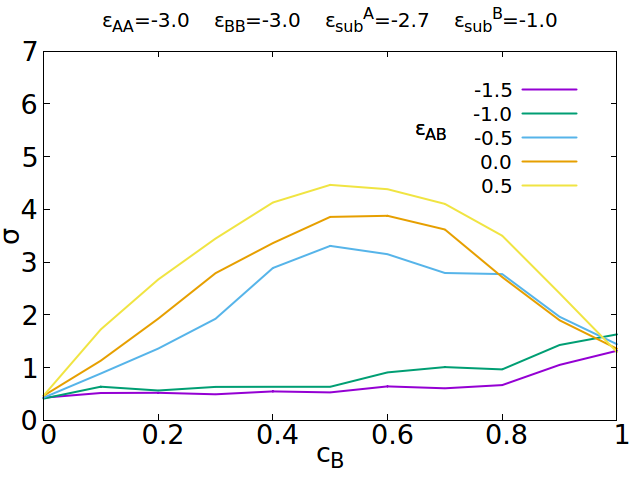}}
  \caption{The effect of interspecies energy variation: In this example, we investigate the demixing case closer and combine a species with island growth with a second that shows 3D growth. This is similar to the C60 mixtures in the main paper.  
	Normalized roughness $\sigma$ as a function of concentration of the island forming species B after deposition of 15 ML. Species A shows 3D growth (left) and LBL (right), variation of $\epsilon_\text{AB}$.  Here, the roughness increase for strong demixing conditions ($\epsilon_\text{AB}=0$ or 0.5) is much more prominent than in the previous case.
It is strongest in the absence of ES barriers (right). An important conclusion is that the experimentally observed decreased roughness in mixed films under (weak) mixing conditions is not seen here. Under demixing conditions the model predicts always an increased roughness whereas in the experiment both
increased and decreased roughness is seen. 
   Other parameters: $\Gamma=10^4$, $E^\text{ES}=3.0$ (left) resp. 0.0 (right), system size $100^2$.}
  \label{fig:eab_2}
\end{figure}

\clearpage

\begin{figure}[h]
  \centerline{\includegraphics[width=18cm]{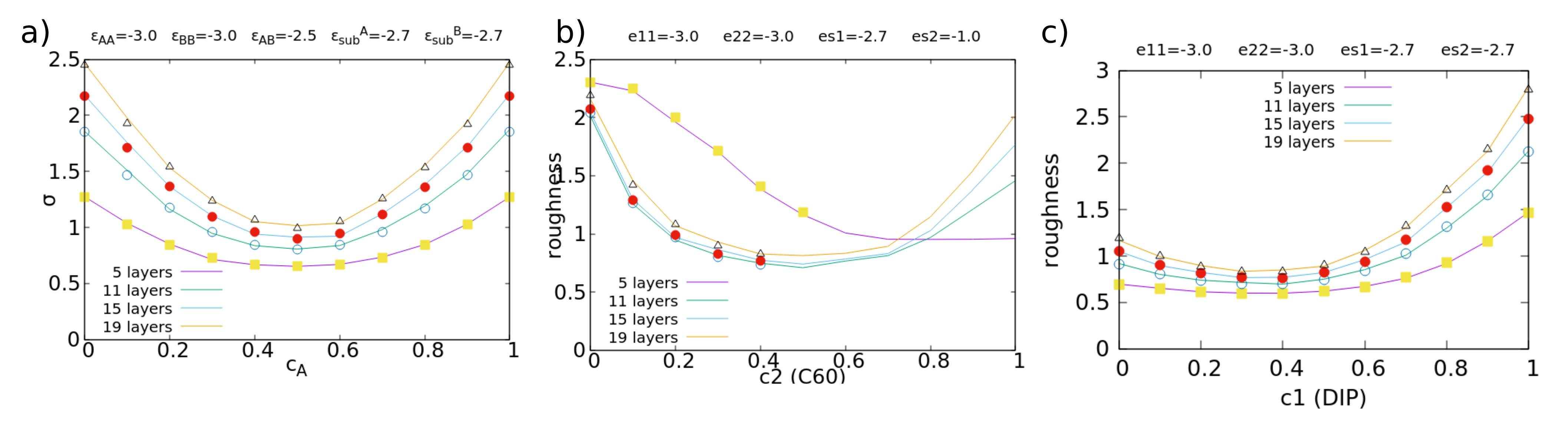}}
  \caption{System size: Comparison of simulation results for different grid sizes: a) PEN:DIP, b) DIP:C60, c) DIP:PDIR. Solid lines correspond to a grid size of $M=300$, symbols correspond to a grid size of $M =800$. We remark that the roughness only weakly depends on the lateral size M of the system. We illustrate this here for 3 examples, where lines are for the small system (M=300) and symbols for the bigger one (M=800). Both system sizes give identical results.}
  \label{fig:M800}
\end{figure}